\documentclass{Interspeech}

\interspeechcameraready

\title{When Humans Growl and Birds Speak: High-Fidelity Voice Conversion \\ from Human to Animal and Designed Sounds}

\author[affiliation={1}]{Minsu}{Kang}
\author[affiliation={1}]{Seolhee}{Lee}
\author[affiliation={1}]{Choonghyeon}{Lee}
\author[affiliation={1,2}]{Namhyun}{Cho}

\affiliation{}{NC AI Co., Ltd}{Republic of Korea}
\affiliation{}{Sogang University}{Republic of Korea}

\email{\{mskang, seolhee, choonghyeon, cnh2769\}@ncsoft.com}

\usepackage{comment}
\usepackage{amsmath, amssymb}
\usepackage{multirow}
\usepackage{subcaption} 
\usepackage{float}
\usepackage{hyperref}

\begin{document}

\maketitle

\begin{abstract}
   Human to non-human voice conversion (H2NH-VC) transforms human speech into animal or designed vocalizations. Unlike prior studies focused on dog-sounds and 16 or 22.05kHz audio transformation, this work addresses a broader range of non-speech sounds, including natural sounds (lion-roars, birdsongs) and designed voice (synthetic growls). To accomodate generation of diverse non-speech sounds and 44.1kHz high-quality audio transformation, we introduce a preprocessing pipeline and an improved CVAE-based H2NH-VC model, both optimized for human and non-human voices. Experimental results showed that the proposed method outperformed baselines in quality, naturalness, and similarity MOS, achieving effective voice conversion across diverse non-human timbres. Demo samples are available at \url{https://nc-ai.github.io/speech/publications/nonhuman-vc/}
\end{abstract}

\newcommand{\minipar}[1]{\smallskip\noindent\textbf{#1}\quad}
\newcommand{\miniparn}[1]{\smallskip\noindent#1\quad}

\section{Introduction}



    Deep-learning based voice conversion (VC) has attained high-fidelity speech transformation \cite{NANSY, DDDM-VC, Diff-HierVC, HierVST}. However, conventional VC methods cannot reliably synthesize non-speech vocalizations demanded by games, sci-fi, and interactive media. Consequently, production pipelines still rely on handcrafted, complex, time-intensive processes that inflate cost and turnaround time. To automate the process, this study explores the human to non-human VC (H2NH-VC) task: given intelligible human vocalizations, the system must generate stylized non-human vocalizations, including screams, growls, dog barks, birdsong, and synthetic orc- and goblin-style utterances.

\begin{figure}[t]
        \centering
        \begin{subfigure}{0.495\columnwidth}
            \centering
            \includegraphics[width=\textwidth]{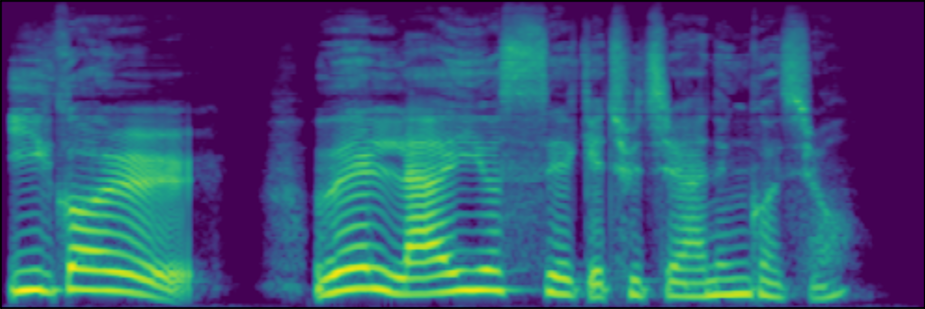}
            \caption{Human Speech}
            \label{fig:human-speech}
        \end{subfigure}
        \begin{subfigure}{0.495\columnwidth}
            \centering
            \includegraphics[width=\textwidth]{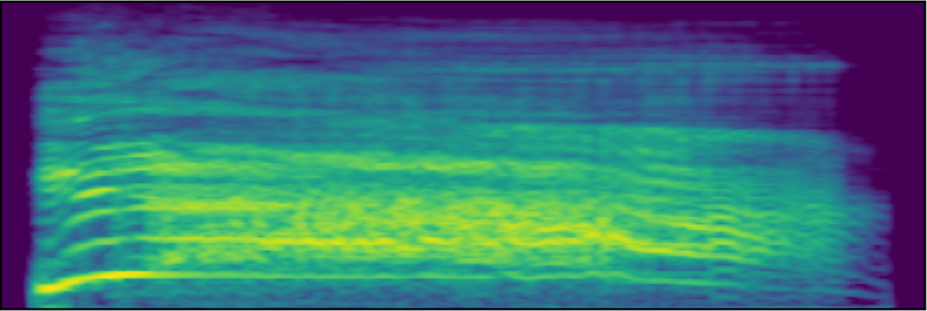}
            \caption{Human Yelling}
            \label{fig:human-yell}
        \end{subfigure}
        
        \begin{subfigure}{0.495\columnwidth}
            \centering
            \includegraphics[width=\textwidth]{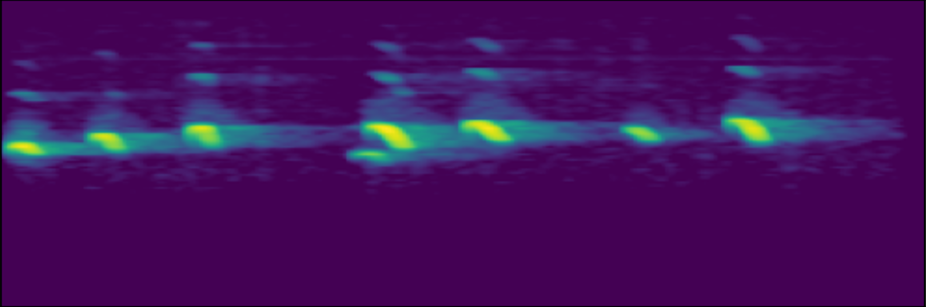}
            \caption{Birdsong}
            \label{fig:non-human-bird}
        \end{subfigure}
        \begin{subfigure}{0.495\columnwidth}
            \centering
            \includegraphics[width=\textwidth]{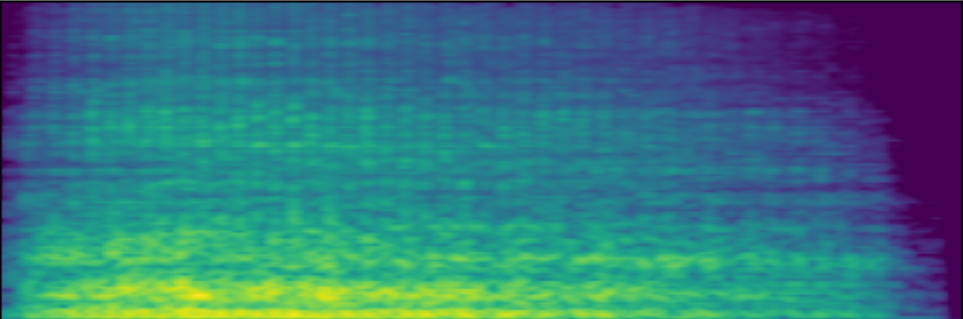}
            \caption{Designed Growling Sound}
            \label{fig:non-human-monster}
        \end{subfigure}
        \caption{mel spectrogram visualizations including examples of (a) human speech, (b) human vocalization: yelling, (c) animal sound: birdsong, (d) artificailly designed sound.}
        \label{fig:human-and-non-human}
    \end{figure}

    Compared to human speech, non-human voices (1) span a broader frequency range and (2) exhibit finer temporal details. For instance, bird-calls occupy higher frequency bands (Fig.~\ref{fig:non-human-bird}), while lion growling exhibits abrupt spectral shifts over extremely short durations (Fig.~\ref{fig:non-human-monster}). Designed non-human voices can vary widely in frequency range and temporal patterns, depending on the types and intensities of effects. The mel spectrograms in Fig.~\ref{fig:human-and-non-human} illustrate these differences: (a) depicts the typical frequency distribution of human speech, (b) depicts high-intensity nonlinguistic expressions (human yelling), (c) depicts bird-call with dense high-frequency components, and (d) depicts a designed growl with pronounced transient characteristics.


    Non-human voices exhibit ambiguous style attributes, making categorization challenging. For instance, the vocal trajectory of an orc in a game is not linked to a specific speaker or emotion, facilitating multiple characters to share the same timbre. While different bird species exhibit timbral differences, identifying individual birds or their emotional variations remains difficult.

    \begin{figure*}[t!]
            \centering
            \begin{subfigure}{0.48\textwidth}
                \centering
                \includegraphics[width=\textwidth]{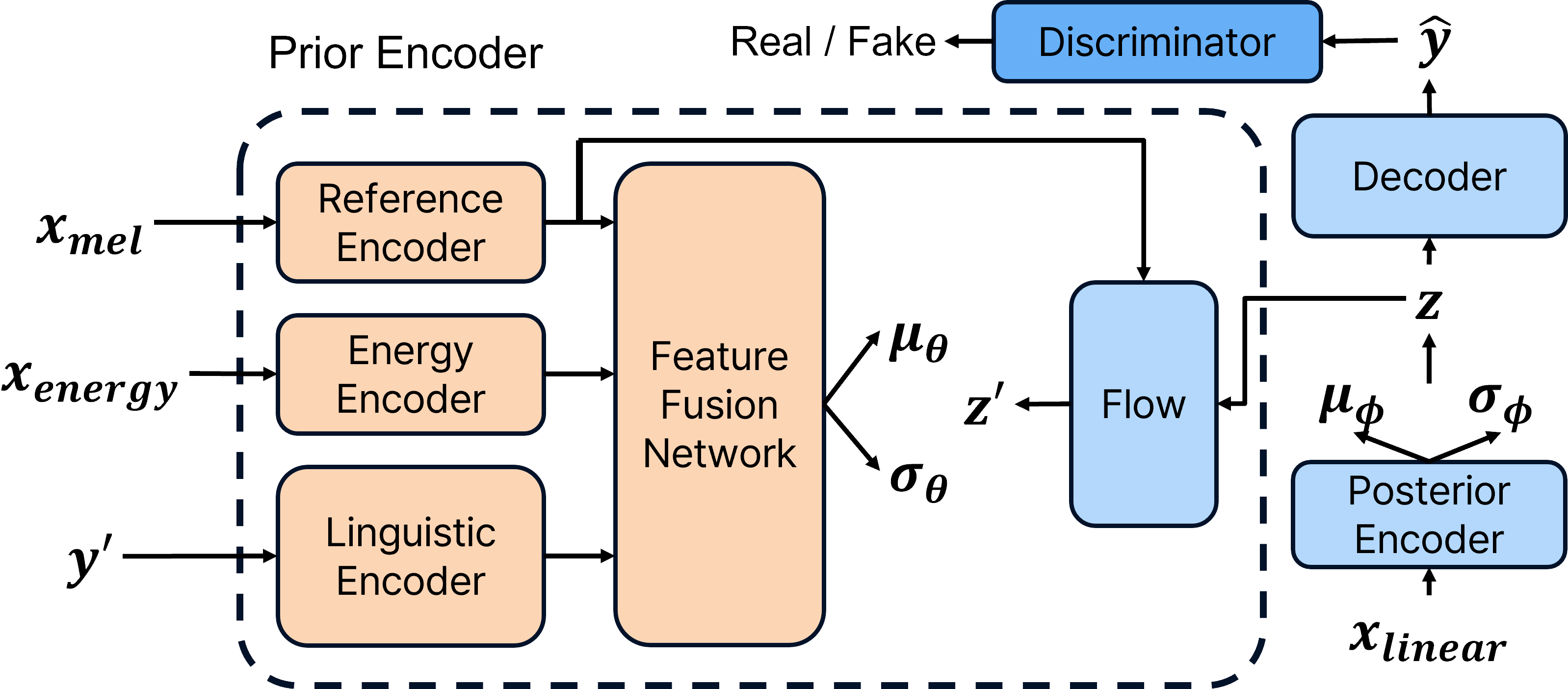}
                \caption{Training}
                \label{fig:model_architecture-training}
            \end{subfigure}
            \hfill
            \begin{subfigure}{0.48\textwidth}
                \centering
                \includegraphics[width=\textwidth]{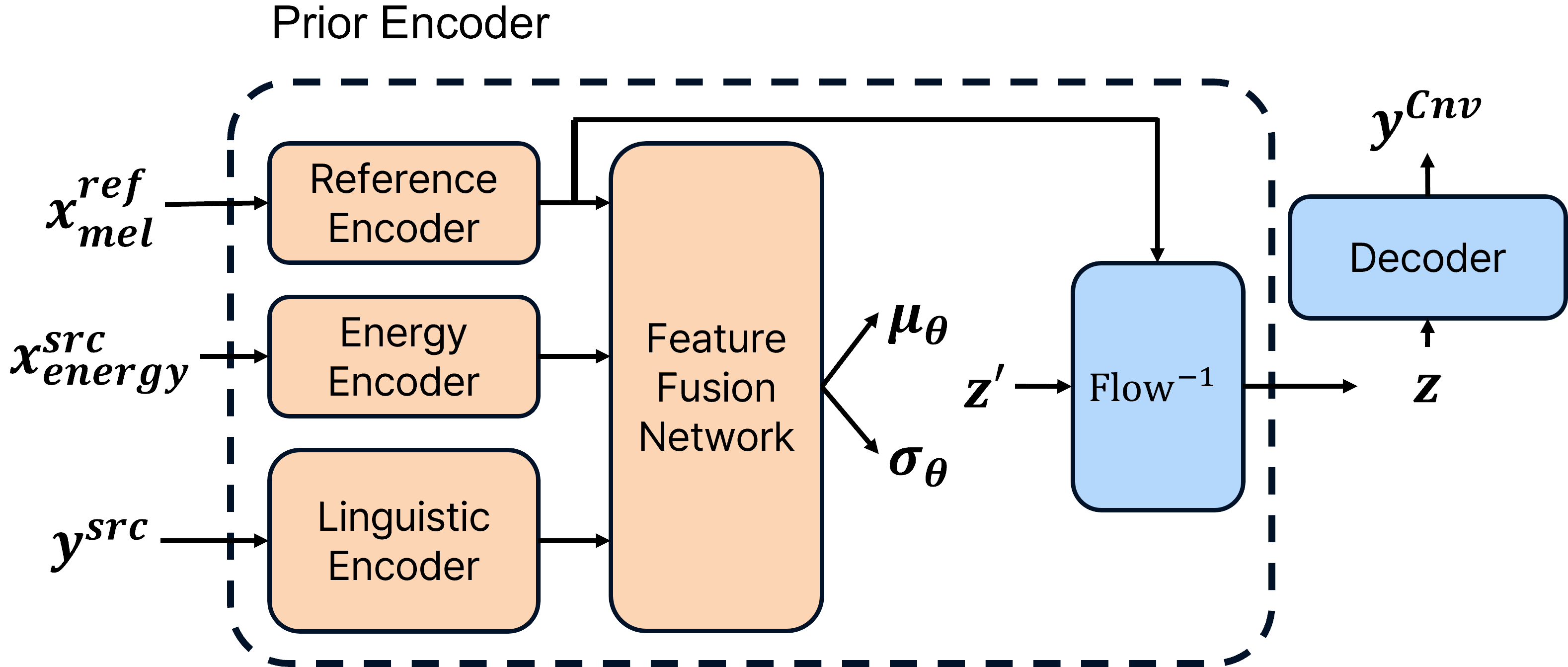}
                \caption{Conversion}
                \label{fig:model_architecture-inference}
            \end{subfigure}
            \caption{Proposed model architecture}
            \label{fig:model_architecture}
        \end{figure*}
    

    Recent human VC studies \cite{NANSY, DDDM-VC, Diff-HierVC, HierVST, Free-VC} have optimized input features for human speech, assuming stationarity by 40-80 frames and 10-20ms hop lengths while primarily modeling the 11.025kHz range with sampling rates of 16 or 22.05kHz. However, non-human voices require broader frequency ranges and finer temporal resolution, rendering human speech preprocessing inadequate. Additionally, most methods incorporate inductive biases specialized for human speech, such as the source-filter theory \cite{NANSY, DDDM-VC, Diff-HierVC, HierVST} or auxiliary losses emphasizing lower frequencies \cite{HierVST}, hindering accurate modeling of non-human sounds.
    

    Speak-Like-a-Dog \cite{Speak-Like-a-Dog} in H2NH-VC attempts voice transformation, but has three key limitations: (1) it relies on human speech preprocessing, failing to capture non-human voice characteristics, (2) it exclusively focuses on dog-sounds, making generalization to other natural or sound-designed non-human voices difficult, and (3) it requires style IDs during training, which are inherently ambiguous in non-human voices, limiting the use of a large-scale non-human dataset.

    
    To address these issues, we propose a methodology for converting diverse non-human voices. We processed audio sampled at 44.1kHz using a Short-Time Fourier Transform (STFT) with a 5ms hop length to generate linear spectrograms, ensuring high fidelity and fine-grained temporal resolution. Mel-filters spanning 0-22.05kHz were applied to capture a broader frequency range. Next, we improved the Conditional adversarial VAE (CVAE)-based VC architecture for H2NH-VC by applying the style vector only to the prior network and flow module, excluding it from the posterior encoder and decoder, enabling richer style representations. Frequency Domain Reconstruction Loss (FDRL) \cite{DAC} was incorporated to recover abrupt transient signals, and KL annealing was applied to mitigate KL-vanishing. Therefore, our model achieved 3.16, 3.16, and 3.78 in quality, naturalness, and similarity MOS, respectively, and reached better in most qualitative metrics, surpassing previous methods.\\
    The main contributions of this study are as follows:
    \begin{itemize}
        \item Unlike prior work \cite{Speak-Like-a-Dog} limited to dog-sound conversion, our approach enables broader non-human voice conversion, including natural and sound-designed voices
        \item We introduce a H2NH-VC model supporting 44.1kHz high-quality audio transformation
        \item We present a preprocessing pipeline optimized for both human and non-human voices
    \end{itemize}


\section{Related Works} 

    \subsection{Voice to Sound Effect Synthesis}
       

        Several studies \cite{DDSP-SFX, T-Foley} have explored voice to sound effect (SFX) generation. However, these approaches were evaluated solely on an SFX database with seven categories, and primarily focused on generating SFX rather than preserving linguistic content. Hence, they are not directly applicable to voice conversion task, where speech content must be retained while timbre is transformed to match a reference signal.

    \subsection{DSP Toolkit based Sound Design}

        DSP toolkits \cite{Dehumaniser2} facilitate the creation of ``designed voices", a form of non-human sound. However, these tools (1) require extensive parameter tuning when applying or combining multiple effects, and (2) exhibit strong interdependencies among parameters, with proprietary internal algorithms making the relationship between settings and output unclear. Similar to DDSP-based approaches \cite{DDSP-SFX}, this study proposes a deep-learning based approach for automated style conversion instead of relying on complex manual adjustments or DSP-based parameter estimation, thereby enhancing efficiency.

    \subsection{Baseline Framework}
        \label{section:related_works-baseline_framework}


        This study adopted the Conditional adversarial VAE (CVAE) framework proposed in \cite{Free-VC}, which builds upon VITS \cite{VITS}. Leveraging core components of VITS \cite{VITS}--a normalizing flow based prior, GAN and VAE based posterior encoder, decoder and discriminator-- this architecture enables end-to-end waveform generation without a separate vocoder.\\
        \textbf{Posterior Encoder, Decoder and Discriminator:} The model received a linear spectrogram $x_{linear}$ of real audio, estimated distribution parameters $(\mu_\phi, \sigma_\phi)$, and modeled the posterior distribution as: $q_\phi(z \mid x_{linear}) = \mathcal{N}(\mu_\phi, \sigma_\phi^2)$. The decoder reconstructed audio (or mel spectrogram) from the latent variable $z$, while the discriminator performed conditional adversarial training, distinguishing real waveforms from generated ones. KL divergence minimization between the posterior and the prior leads to the alignment of the latent space. \\
        \textbf{Prior Encoder:} The prior encoder received a reference mel spectrogram $x_{mel}$ and timbre-perturbed waveform ($y'$), extracting a style vector ($s$) and linguistic representation ($x_{ling}$). A feature fusion network\footnote{referred to a linguistic restorer in HierVST \cite{HierVST} and the bottleneck extractor in Free-VC \cite{Free-VC}.} computed the prior distribution parameters $(\mu_\theta, \sigma_\theta)$. The normalizing flow module $f(\cdot)$ transformed latent variable $z'$ sampled from a prior distribution $N(\mu_\theta, \sigma_\theta)$ to match complex posterior distribution. Specifically, the transformation followed $z' = f(z)$ during training, whereas inference was performed using $z = f^{-1}(z')$, yielding $p_\theta(z\mid x_{ling}, s) = \mathcal{N}\bigl(z'; \mu_\theta, \sigma_\theta^2\bigr)\left|\det\frac{\partial f^{-1}(z')}{\partial z}\right|$. \\
        \textbf{Training and Inference:} In training, KL divergence loss $D_{KL}=\text{KL}\left(q_\phi(z \mid x_{linear}) \,\Vert\, p_\theta(z \mid x_{ling}, s)\right)$, waveform reconstruction loss $L_{rec}$, adversarial loss $L_{adv}$, and feature matching loss $L_{fm}$, were minimized. In inference, the model extracted style $s^{ref}$ from a reference waveform and linguistic representation $x_{ling}$ from a source waveform. The prior network and flow ($f$) generated latent variable $z$, which the decoder used to synthesize the converted waveform $y^{cnv}$.

        The CVAE-VC framework offers three key advantages: (1) end-to-end waveform generation enables high-quality audio synthesis \cite{VITS, Wave-Tacotron, AM-Voc-JT-Singing}, (2) learning latent acoustic tokens from the VAE encoder, rather than relying on a fixed input representation (e.g., mel spectrograms), enhances efficiency and effectiveness in diverse sound generation \cite{AudioLDM}, and (3) a learnable prior network allows more complex and flexible distribution compared to a fixed and simple normal distribution \cite{HierVST, VITS, VampPrior, Flowtron}.

\section{Proposed Method}
    \label{sec:proposed_method}
    

        This section outlines the preprocessing and feature extraction techniques used to accommodate human and non-human voices and improvements in the model architecture and training methodology. Figure \ref{fig:model_architecture} presents an overview of the workflow of the system.
    
    \subsection{Input Feature Set and Preprocessing Method}

    
        \minipar{Waveform, STFT and Mel:} A sampling rate of 44.1 kHz was employed to ensure high-fidelity audio. STFT was applied using a frame/window length and hop length of 20 ms and 5 ms, respectively, to capture fine-grained temporal details, particularly in non-human sounds. The model learned a mel spectrogram spanning a 0-22.05 kHz range, enabling the incorporation of a broader frequency spectrum.


        \minipar{Style Representation:} A reference encoder extracted style representation directly from the reference audio, instead of relying on predefined style labels. The reference encoder processed an utterance-level mel spectrogram, averaging it along the time axis. This generated a global style vector that captured the overall timbral characteristics of both human and non-human voices. The reference encoder was trained end-to-end alongside the model. The architecture followed the mel-style encoder framework proposed in \cite{Meta-StyleSpeech}.


        \minipar{Linguistic Representation:} Conventional automatic speech recognition based phoneme posteriorgram extractors are inadequate for nonlinguistic or sound-designed voices, since they were trained on speech databases with clearly defined phonetic structures. To learn both phonemic and continuous features independent of timbre information, we use self-supervised learning(SSL) based information perturbation \cite{NANSY, DDDM-VC, Diff-HierVC, HierVST}. Specifically, to accommodate the wide frequency variations in non-human sounds, formant shift, pitch shift, and pitch range parameters were expanded from (1.4, 1.5, 2) to (1.8, 3, 2), generating timbre-perturbed waveforms. The 12th-layer hidden representation of pretrained XLS-R \cite{XLS-R}, a multilingual variant of Wav2Vec2.0 \cite{Wav2vec2.0}, was extracted from 16 kHz inputs with a 25 ms window and 20 ms stride, followed by an interpolation to achieve a 5 ms hop resolution.

        
        \minipar{Prosodic Feature:} Frame-level energy was utilized to preserve the local prosody of the original audio. Frame-level energy was computed as the L2 norm of the linear STFT spectrogram frames, using a 20ms window and 5ms hop length. Log-mean subtraction $y = \log(x+\epsilon) - \log(\mu+\epsilon)$ was applied to maintain the global style vector consistent while preserving local prosody. Additionally, we attempted to utilize frame-level f0 from non-human sounds. However, existing f0 estimation methods \cite{Praat, YAPPT, CREPE, SPICE, PESTO} exhibited limitations due to the absence of a well-defined harmonic structure in non-human sounds. Consequently, this study relied solely on energy features, leaving robust f0 extraction as a direction for future research.
        

    \subsection{Model Architecture and Improvements}

        As shown in Fig.~\ref{fig:model_architecture}, the proposed model consists of a \textit{CVAE}-based posterior encoder, Decoder, prior encoder, and flow module. While the posterior encoder and decoder follow earlier works \cite{HierVST, Free-VC, VITS}, we make two enhancements: (1) we input a 5\,ms–hop linear spectrogram $x_{linear}$ to achieve finer temporal resolution, and (2) we apply the style vector only to the prior network and Flow module, ensuring it does not redundantly appear in the posterior encoder or Decoder. This design avoids style overlap that could degrade performance.

        As illustrated in Fig.~\ref{fig:model_architecture}, the proposed model comprises a CVAE-based posterior encoder, decoder, prior encoder, and flow module. While the posterior encoder and decoder are based on previous architectures \cite{HierVST, Free-VC, VITS}, two key modifications are introduced: (1) A 5ms hop length linear spectrogram $x_{linear}$ was used as input to achieve finer temporal resolution. (2) The style vector was applied exclusively to the prior network and flow module, and removed from the posterior encoder or decoder. This design reduces the risk of style overlap between the linear spectrogram and the style vector, preventing potential degradation in style conversion.


        The prior encoder consisted of a reference encoder, an energy encoder, and a linguistic encoder. The reference encoder extracted style embeddings from a mel spectrogram, while the energy encoder processed 5 ms frame-level energy values using a 1D convolutional neural network (CNN). The linguistic encoder employed the pretrained Wav2Vec2.0 model \cite{Wav2vec2.0} to derive SSL-based features from a 16kHz downsampled waveform. These features were then interpolated along time axis and projected to the dimensionality of $D_{ling}$ at 5ms intervals.
        The representations from these three encoders were concatenated and passed through a single-layer convolutional feature fusion network, which estimated the prior distribution parameters ($\mu_\theta$, $\sigma_\theta$). The sampled latent variable $z'$ was subsequently processed through the flow module $f$, yielding the final latent representation $z$.

        \begin{table*}[t!]
            \caption{Comparison of baseline methods, and ablation study results. "Proposed" denotes our proposed model. “w/o PP” indicates the proposed model using conventional (speech-focused) preprocessing method instead of our proposed pipeline, “w/ SEED” denotes proposed model with style embedding added to the audio encoder/decoder, and “w/o KL-A” represents the proposed model trained without KL Annealing. "Time" refers to the generation time per sample for a 6-second waveform, measured using an A100 GPU.}
            \label{tab:comparison}
            \centering
            \begin{tabular}{lcccccccccc}
                \toprule
                \textbf{Method} 
                & \textbf{MOS-S ($\uparrow$)} 
                & \textbf{MOS-Q ($\uparrow$)} 
                & \textbf{MOS-N ($\uparrow$)} 
                & \textbf{PCC ($\uparrow$)} 
                & \textbf{RMSE ($\downarrow$)} 
                & \textbf{CER ($\downarrow$)} 
                & \textbf{WER ($\downarrow$)}
                & \textbf{Time ($\downarrow$)} \\
                \midrule
                DDDM-VC ~\cite{DDDM-VC} & 3.06 & 2.31 & 2.62 & 0.71 & 0.0661  & 43.24\% & 64.21\% & 2.47 \\
                Diff-HierVC ~\cite{Diff-HierVC} & 3.11 & 2.31 & 2.59 & 0.75 & 0.0632 & 48.61\% & 66.07\% & 0.52 \\
                Free-VC ~\cite{Free-VC} & 2.96 & 2.73 & 3.22 & 0.81 & 0.0442 & 30.97\% & 53.62\% & 0.14 \\
                \midrule
                \textbf{Proposed} & \textbf{3.78} & 3.16 & 3.16 & \textbf{0.99} & \textbf{0.0326} & 15.48\% & 24.02\% & 0.12 \\
                  \midrule
                   w/o PP & 3.53 & \textbf{3.30} & 2.96 & 0.80 & 0.0345 & 42.44\% & 62.29\% & 0.09 \\
                   w/ SEED          & 3.61 & 3.19 & \textbf{3.28} & 0.98 & 0.0352 & \textbf{10.63\%} & \textbf{19.29\%}  & -- \\
                    w/o KL-A          & 3.77 & 3.04 & 3.17 & \textbf{0.99} & \textbf{0.0326} & 28.89\% & 44.69\% & -- \\
                \bottomrule
            \end{tabular}
        \end{table*}

    \subsection{Training}

        The total loss function comprised generator ($L_{gen}$) and discriminator ($L_{dis}$) losses. The generator loss is defined as a weighted sum of the KL divergence loss $L_{KL}$, the feature matching loss $L_{fm}$ \cite{Feature-Matching-Loss}, the LS-GAN based adversarial loss $L_{adv}$ \cite{LS-GAN}, and the reconstruction loss $L_{rec}$. In \cite{HierVST, Free-VC, VITS}, the fixed-hop-length Mel-STFT loss can struggle to reconstruct non-human audio with significant transient variations. To mitigate this, we use FDRL \cite{DAC}, which employs multiple short and long hop lengths, effectively captures fine-grained temporal fluctuations \cite{DAC}. A DAC discriminator \cite{DAC} was incorporated, and trained using the adversarial loss, defined as $L_{dis}=L_{adv}(D)$ \cite{LS-GAN}.


        \minipar{Cosine KL Annealing:} Previous studies \cite{HierVST, Free-VC, VITS} apply a fixed weight to the KL divergence loss(i.e., $\lambda_{KL}=1$). However, this can over-regularize an underdeveloped prior in early training stages, leading to KL-vanishing (Posterior Collapse) \cite{PosteriorCollapse2}. Hence, a cosine KL annealing scheme (Eq.~\ref{Equation:Cosine-KL-annealing}) was employed, wherein $\lambda_{KL}$ gradually increased from near zero to one over $t_{anneal}$ steps. This approach alleviated KL pressure in the initial training phase while progressively enforcing stronger prior–posterior alignments as training advances.
        \begin{equation}
            \lambda_{KL} = \min\left(\frac{1}{2}\cos\left(\pi\left(\frac{t_{cur}}{t_{anneal}}-1\right)+1\right), 1\right)
            \label{Equation:Cosine-KL-annealing}
        \end{equation}
            


\section{Experiments}
    
    \subsection{Experimental Settings}
        We used internally collected dataset consisted of 82,008 audio samples, including the CORE 6 library from Pro Sound Effects \cite{Core6Library}. The dataset was categorized as follows: (1) Exclamations and expressive utterances (37,332 samples), encompassing vocal shouts, laughter, and similar expressions. (2) Sound-designed non-human voices (42,800 samples), representing synthesized characters such as goblins and zombies. (3) Animal sounds (1,886 samples), including recordings of lions, birds, and other species. All audio samples were stored in a 44.1 kHz, 16-bit WAV format. 

        %
        %


        The proposed model architecture was configured as follows: The reference encoder utilized 128 hidden dimensions, while the posterior encoder, decoder, flow, and feature fusion networks each employed 192 hidden dimensions. The decoder applied an upsampling rate of (11, 5, 2, 2) with an initial channel size of 1024. The FDRL \cite{DAC} was computed using hop lengths of (882, 441, 220, 110, 55), with window lengths set to four for each hop, and 80 mel bins. The annealing step $t_{anneal}$ set to 50,000, and loss weights set as follows: $\lambda_{rec} = 45$, $\lambda_{fm} = 2$, and $\lambda_{adv} = 1$. The model was trained using four NVIDIA A100 GPUs with a batch size of 128, employing the Adam-W optimizer for up to 400,000 steps. Each input segment comprised 100 frames.


        The evaluation involved both subjective and objective assessments. Subjective evaluation: 5-scale MOS tests were conducted with 9 participants to assess quality (MOS-Q), naturalness and pronunciation clarity (MOS-N), and similarity (MOS-S). Objective metrics: (1) PCC-E: Pearson correlation coefficients between source and converted energy contours. (2) RMSE-E: Root mean squared errors between energy values of the source and converted signals. (3) CER/WER: Assessed using Whisper \cite{Whisper} solely on samples that contain lingual phonemic information.

        
        
        
        
        The proposed model was compared against the following baselines: (1) DDDM-VC \cite{DDDM-VC}: a VC model integrating a source-filter–based diffusion acoustic model with a separate vocoder \cite{HiFi-GAN}, designed to enhance the disentanglement of prosodic elements in human speech. (2) DiffHierVC \cite{Diff-HierVC}: a diffusion-based model employing a hierarchical structure to more accurately capture the source-filter properties of human speech over time and frequency. A separate vocoder \cite{HiFi-GAN} was used for waveform reconstruction. (3) Free-VC \cite{Free-VC}: a CVAE-VC model implementing text-free VC, focusing on detailed linguistic representation extraction. All baseline trained using conventional speech-focused pipelines.

            
            

    \subsection{Experimental Results}
    
        \subsubsection{Comparison of baseline methods: }
            \label{section:comparison_of_baseline_methods}
            Table~\ref{tab:comparison} shows that our proposed model outperforms the baselines in most objective and subjective metrics. Specifically, our model achieves a PCC-E of 0.99, and an RMSE-E of 0.03256, along with significantly lower CER 15.48\% and WER 24.02\%. Subjective evaluations indicate MOS-Q and MOS-S scores of 3.16 and 3.78, respectively, which are higher than those of all baselines. Notably, Free-VC \cite{Free-VC} obtains a MOS-N score of 3.22, which is slightly higher than our model. This may be because the model, having lower conversion performance, tends to preserve expressive details in the source audio, making the converted sound more natural to participants.
            
        \subsubsection{Effect of Preprocessing Settings: }
            Replacing our proposed preprocessing pipeline with a conventional speech-focused approach (w/o PP) led to a noticeable drop in similarity (MOS-S: 3.78→3.53) and naturalness (MOS-N: 3.16→2.96), showing that the converted voices were perceived as less similar and natural to the reference. From the objective metrics, PCC-E decreased from 0.99 to 0.80, and RMSE-E slightly increased from 0.03256 to 0.03450. This suggests that while the overall energy patterns were still relatively well-preserved, the conventional preprocessing struggled to capture fine transients, which are crucial for non-human sound characteristics. Furthermore, CER increased dramatically (15.48\%→42.44\%) and WER also worsened significantly (24.02\%→62.29\%), indicating that linguistic content was much less accurately retained. These results emphasize the effectiveness of our proposed preprocessing method in maintaining both the detailed time-frequency structures of non-human sounds and the clarity of linguistic information.
            
        \subsubsection{Ablation Study: }
                       
            \textbf{Effect of Adding Style Emb. in the Audio Enc./Dec.:} Adding the style embedding (SEED) to the audio encoder and decoder reduced similarity (MOS-S: 3.78→3.61), likely due to style overlap between the style embedding from reference encoder and latent acoustic tokens from acoustic encoder. PCC-E and RMSE-E are almost same as (0.99→0.98, 0.03256→0.03519). However, similar to the trend observed in comparison of baseline methods(\ref{section:comparison_of_baseline_methods}), a tradeoff emerged between similarity and quality/naturalness, where the addition of SEED improved MOS-Q and MOS-N (3.16→3.19 and 3.16→3.28, respectively) at the cost of reduced similarity.\\
            \textbf{Effect of Removing KL Annealing (w/o KL):} Without KL annealing, MOS-Q dropped from 3.16 to 3.04, while MOS-N and MOS-S remained nearly the same(3.16→3.17, and 3.78→3.77). However, a substantial increase in CER (15.48\%→28.89\%) and WER (24.02\%→44.69\%) suggests that removing KL annealing weakened latent space regularization, making it more difficult to maintain clear and expressive linguistic content in the converted speech.

\section{Conclusion}
    The proposed H2NH-VC model effectively converted human speech into diverse non-human timbres, achieving significant MOS (Q/N/S) improvements over prior methods. It accurately captured transient-rich signals and wide-frequency–range of non-human sounds. The proposed method outcomes hint at its strong potential for use cases demanding extreme timbral changes—such as game audio or sci-fi film sound design—where style consistency and control are paramount.

\section{Acknowledgements}
This research was supported by Culture, Sports and Tourism
R\&D Program through the Korea Creative Content Agency
grant funded by the Ministry of Culture, Sports and Tourism
in 2024 (Project Name: Development of Co-Pilot technology
for automatic completion of generative AI-based 3D Webtoon,
Project Number: RS-2024-00400004, Contribution Rate: 100\%)

\bibliographystyle{IEEEtran}
\bibliography{mybib}

\end{document}